\documentclass[11pt]{article}

\usepackage{latexsym}
\usepackage{amsfonts}
\usepackage[usenames,dvipsnames]{xcolor}
\usepackage[english]{babel}
\usepackage{graphicx}
\usepackage{dcolumn}
\usepackage{cancel}
\usepackage{float}
\usepackage{amsmath}
\usepackage{amsthm}
\usepackage{amsfonts}
\usepackage{amssymb}
\usepackage{booktabs}
\usepackage{color}
\definecolor{darkgreen}{rgb}{0,0.5,0}
\definecolor{purple}{rgb}{1,0,1}
\newcount\Comments  %
\newcommand{\kibitz}[2]{\ifnum\Comments=1\textcolor{#1}{#2}\fi}

\usepackage{srcltx}
\usepackage{authblk}

\usepackage{setspace}
\definecolor{myblue}{rgb}{0,0,0.8}

\definecolor{green}{RGB}{0, 130, 0}
\definecolor{grey}{RGB}{90, 90, 90}

\catcode`\@=11
\def\marginnote#1{}

\topmargin by -\headsep
\def\numberbysection{\@addtoreset{equation}{section}
        \def\theequation{\thesection.\arabic{equation}}}

\def\underline#1{\relax\ifmmode\@@underline#1\else
 $\@@underline{\hbox{#1}}$\relax\fi}

\catcode`@=12 \relax

\numberbysection

\def\nonu{\nonumber}
\def\br{\begin{eqnarray}}
\def\er{\end{eqnarray}}
\def\lb{\lbrack}
\def\rb{\rbrack}

\def\({\left(}
\def\){\right)}
\def\[{\left[}
\def\]{\right]}

\def\lie{{\cal G}}
\def\a{\alpha}

\def\eps{\epsilon}

\def\vareps{\varepsilon}

\def\pa{\partial}

\def\tp0{\Theta_{+}^{(0)}}
\def\tm0{\Theta_{-}^{(0)}}

\def\nonu{\nonumber}

\begin{document}


\title{Generalized Riemann-Hilbert-Birkhoff Decomposition  and a 
 New Class of Higher Grading Integrable Hierarchies}

\author[1]{H. Aratyn }
\author[2]{C.P. Constantinidis }
\author[3]{J.F. Gomes}
\author[3]{T.C. Santiago}
\author[3]{A.H. Zimerman}

\affil[1]{
Department of Physics, 
University of Illinois at Chicago, 
845 W. Taylor St.
Chicago, IL 60607-7059, USA}
\affil[2]{
Universidade Federal do Espirito Santo,
Depto. de F\'{\i}sica, 
Av. Fernando Ferrari, 514,
UFES/DFIS 29075-900 Vitoria-ES, Brazil}
\affil[3]{
Instituto de F\'{\i}sica Te\'{o}rica-UNESP,
Rua Dr Bento Teobaldo Ferraz 271, Bloco II,
01140-070 S\~{a}o Paulo, Brazil
}

\maketitle

\abstract{
We propose a generalized Riemann-Hilbert-Birkhoff decomposition that
expands the standard integrable hierarchy formalism in two fundamental
ways: it allows for integer powers of Lax matrix components in the
flow equations to be increased as compared to conventional models, 
and it incorporates
constant non-zero vacuum (background) solutions.

 Two additional parameters control these features.
The first one defines the grade of a semisimple element that underpins
the algebraic construction of the hierarchy, where a
grade-one semi-simple element recovers known hierarchies such as mKdV
and AKNS. The second parameter distinguishes between zero and non-zero constant 
background (vacuum) configurations.

Additionally, we introduce a third parameter associated with an
ambiguity in the definition of the grade-zero component of the
dressing matrices. While not affecting the decomposition itself, this
parameter classifies different gauge realizations of the integrable
equations (like for example, Kaup-Newell, Gerdjikov-Ivanov, Chen-Lee-Liu models).

For various values of these parameters, we construct and analyze
corresponding integrable models in a unified  universal manner  demonstrating the
broad applicability and generative power of the extended formalism.
}

\section{Introduction}

The algebraic formalism  offers a powerful framework to
construct and to classify integrable hierarchies.  It rests
on a  zero curvature relations solved by  Lie algebra valued gauge potentials 
(Lax operators) obtained via recursive considerations  \cite{drinfeld-sokolov}, \cite{leznov-saveliev}, \cite{olive},\cite{mira2}, \cite{adler}.
A direct consequence of these relations is the existence of infinite
number of conservation laws that are responsible for presence and 
stability of soliton solutions.  Soliton solutions 
are  very special solutions that
preserve their form and interact elastically with other solitons.  In
general, non-trivial solitons are constructed from a basic zero vacuum solution by
gauge transformation (dressing transformation) \cite{babelon}, e.g. 
Korteweg-de Vries (KdV) and  modified Korteweg-de Vries (mKdV)
equations.  Interestingly,  there is also a class
of non-linear equations, that only admits strictly
non-zero constant vacuum solutions (see for instance \cite{gui1}).
This fact defines two distinct orbits, zero and non-zero constant vacuum,
that will be embodied into a single  formalism we shall  propose.

It is well-known that construction of commutative (positive and
negative) flows satisfying zero curvature relations follows straightforwardly 
from the Riemann-Hilbert-Birkhoff (RHB) formalism \cite{RH,RH2}. It is based upon   an affine Lie algebraic structure
with decomposition into graded subspaces induced by a grading operator 
$Q$ (see for instance \cite{AGZ} ). 
{In this setting we explore various ambiguities related to
formulation of the underlying Heisenberg algebra in  the graded
environment. One ambiguity is connected with a freedom to define the 
decomposition formalism in terms of every second grade or 
every third grade and so on, rather than the conventional 
every neighboring grade 
(see discussion below equation 
\eqref{psizero} for explicit realization and discussion of this idea). 
Additionally there also exists an ambiguity that accounts for 
freedom of adding extra terms
to the semisimple element generating the flows in the RHB formalism
without a change to the Heisenberg algebra it satisfies but giving
raise to the non-zero constant vacuum solution.
 }
Recognizing these ambiguities we here  generalize  the RHB decomposition 
by augmenting it with additional structures that are
controlled by new  $a$ and $b$ parameters. The  parameter $a$ enters
directly 
the definition of a semi-simple element that is a starting block of
the decomposition formalism  and effectively
increases the powers of fields entering the flow equations while $b$ 
switches  on and off the non-zero vacuum solution. The models with $a>1$ will be
referred to as higher grading hierarchies. We will refer to the generalized
Riemann-Hilbert-Birkhoff decomposition by the abbreviation 
g-RHB.
  
  In general, solitons are constructed by applying dressing
  transformation to  the zero vacuum
  solution \cite{babelon,mira}.  In practice this is accomplished algebraically by
gauge transforming the two dimensional gauge potential
  from the vacuum configuration to some non-trivial configuration.  
  The key point in
  our observation is that some hierarchies present more than one type
  of vacuum solution, i.e.  zero or constant non-zero vacuum, e.g.
  mKdV , Chen-Lee-Liu model, etc. 
  Each type of such solution is characterized by a
  Heisenberg algebra underlying the two dimensional gauge potential in
  vacuum, each having its own  different Heisenberg
  algebra.  The Heisenberg algebra, in turn, depends upon the vacuum
  parameters controlled here by a parameter $b$.  The soliton
  solutions are therefore classified by orbits generated from 
  vacuum solutions and characterized by their own  different Heisenberg algebras.
  Related to the presence of the $b$ parameter it is worthwhile to
  point out that in the reference \cite{gui1} the negative even-graded
  sub-hierarchy of mKdV was proposed and its solitons were constructed
  from a strictly non-zero vacuum solution.  This was later extended
  to the KdV hierarchy by Miura transformation and an interesting
  degeneracy of zero and non-zero vacuum solutions was pointed out in
  \cite{jhep-paper}.  

In addition, we
also employ a new parameter $c$ that, 
reflects ambiguity of the  zero-grade
components of the dressing matrices and realizes different gauge 
formulations of the underlying flow equations. Gauge equivalence via
Miura transformations connecting different splitting of subalgebras 
associated with the zero grade ambiguity has been observed in the KP
hierarchy treated by Adler-Kostant-Symes $R$-matrix scheme \cite{ANPV}.

 In this paper we show that  Kaup-Newell, Gerdjikov-Ivanov 
  and  Chen-Lee-Liu hierarchies, respectively, 
 can be derived directly from  the  g-RHB formalism 
 for the value of the parameter  $a$ equal to $a=2$ and 
 for the appropriate values of the parameter $c$.  
 Although their flow equations  are related to each other through 
 gauge ambiguity 
 classified by the parameter $c$,  remarkably the Chen-Lee-Liu  hierarchy
 is the only one that admits non-zero constant  vacuum solutions.
In the related forthcoming
publication 
\cite{thais-paper} the phenomenon of strictly non-zero vacuum
will be shown to appear within the Chen-Lee-Liu hierarchy.  In particular in  
\cite{thais-paper} it will be shown that the constant non-zero solitons
solutions can be employed systematically in order to reduce the Chen-Lee-Liu 
hierarchy into the Burgers hierarchy.

The g-RHB formalism allows for a general construction of  integrable 
hierarchies with both homogeneous and principal 
gradations but here we will focus on producing higher grading
hierarchies models with principal  gradation  for which we are also
able to study  the new types of vacuum structures. 

The paper is organized as follows. 

The background material is presented in section \ref{section:algebra}, 
where we show how zero curvature relations  are employed within algebraic
approach
to derive the commuting time evolution
equations for both positive and negative flows.
In section \ref{section:GRHB}, the algebraic construction of 
section \ref{section:algebra} is embedded in a formalism
 of 
 g-RHB decomposition
 of an affine Lie algebra.  This    setting  allows generalizations
 to ``higher grading'' integrable hierarchies in a  unified framework that is
 controlled by the three parameters $a,b$ and $c$, which we introduce 
 to describe various  models and their realizations.
 Section \ref{section:models}  presents integrable hierarchies obtained 
by construction presented in Section  \ref{section:GRHB}.

The models that are characterized by principal gradation  and $a=1,2$ are
presented in subsections \ref{subsection:mkdv} and \ref{subsection:kn}.
The higher gradation model with $a=3$ is given in section
\ref{section:higher}.
These results are summarized in the below table that lists
principal gradation models obtained 
from the g-RHB decomposition
formula \eqref{gr-h} corresponding to the choice of algebra $\hat {\lie}$ (here
fixed to be $\hat{sl}_2$), 
and the values of the parameters $a$ and $c$.


\begin{table}[h]
\caption{Models obtained from g-RHB decomposition with $\hat {\lie}=\hat{sl}_2$ and principal
gradation}
    \setlength{\tabcolsep}{4pt}
\begin{tabular*}{\textwidth}{@{\extracolsep\fill}|c|l|l|}\hline
 $a$  & c & Hierarchy\\  \hline
 $a=1$ &$0\leq c \le 1$ & mKdV
\\ \hline
 $a=2$ &$c=0$ & GI\\ \hline
 $a=2$ &$c=1/2$ & CLL\\ \hline
 $a=2$ &$c=1$ & KN\\ \hline
 $a=3$ &$0\leq c \le 1$ & $v_1,v_2,v_3$ hierarchy of section
\ref{section:higher} \\ \hline
\end{tabular*}
\end{table}
In the above table we have used the following well-known integrable
equations denoting them by their
well-established abbreviations: 
modified Korteweg-de Vries (mKdV), 
Kaup-Newel (KN), Chen-Lee-Liu (CLL) and  Gerdjikov-Ivanov (GI) that
will be used consistently throughout the rest of this paper. 

In addition, it is worthwhile to point out that the well-known 
homogeneous $\hat{sl}_2$ hierarchies 
Ablowitz-Kaup-Newell-Segur (AKNS)  and its connection  to  Wadati-Konno-Ishikawa (WKI) hierarchy (see \cite{jhep-paper}) corresponds 
to $a=1$,  $c=0$ and $c=1$  respectively,   can be  reproduced by g-RHB formalism as will be explained in a separate publication.

\section{Algebraic Background}
\label{section:algebra}
The algebraic approach to studying integrable hierarchies forms 
a leading framework for systematically formulating hierarchies of time 
evolution (flow) equations, their infinite conservation laws, 
and for methodically deriving the corresponding soliton solutions.
In order to develop such approach one starts with a  
decomposition of affine  Lie algebra $\hat {\lie}$ 
 into graded subspaces 
$ \hat{\mathcal{G}}=\oplus_{i \in \mathbb{Z} }{\mathcal{G}}_i$,
with   ${\mathcal{G}}_i$ {satisfying} 
\br
    \lb {Q}, {\mathcal{G}}_i\rb =i {\mathcal{G}}_i,
    \qquad  \lb {\mathcal{G}}_i, {\mathcal{G}}_j\rb 
    \subset {\mathcal{G}}_{i+j}, \label{g1}
\er
where $Q$ is a grading operator { and $i,j \in \mathbb{Z}$}.

For a
semi-simple element of grade $a$, $E \equiv  E^{(a)} \in
{\mathcal{G}}_a$  an affine  Lie algebra $\hat {\lie}$
further decomposes into the kernel of $E$ 
defined  by 
\begin{equation*} \mathcal{K}\equiv \left\{ X \in
\hat{\mathcal{G}} \,|\, \lb{E},{X}\rb =0\right\}, \end{equation*} 
and its complement $\mathcal{M}$ :
\begin{equation*}
\hat{\mathcal{G}}=\mathcal{K}\oplus \mathcal{M}\, .\end{equation*}
The integrable hierarchies are henceforth classified according to the 
decomposition of the affine algebra $\hat {\lie} $ in terms 
of a grading operator $Q$ and  a constant 
semi-simple generator $E \in \hat {\lie}$.

{Within such algebraic structure it is convenient to introduce  
zero curvature relations in order  to derive the commuting time evolution
equations.  A simple example of hierarchy defined in terms of
a semisimple element  of grade one, $E^{(1)} \in
{\mathcal{G}}_1$
will illustrate basic features that will be easy to recognize 
in a variety of models presented below.  The evolution flows defining
such hierarchy 
are derived from the  zero curvature relations in terms of  a two dimensional gauge potentials, namely,  $A_x$ and $A_{t_N}$:
\br
 [ \mathcal{D}_x + A_x, \;  \mathcal{D}_{t_{N}}+ A_{t_N}]= [  \mathcal{D}_x + E^{(1)} + A_0, \;  \mathcal{D}_{t_N}+D^{(N)}  + D^{(N -1)}\cdots + D^{(0)} ]=0, 
\label{1}
\er
$ E^{(1)} \in \lie_1, \; A_0 \in \lie_0, \;  { N \in \mathbb{Z}_+^{*}}$ and $\mathcal{D}_{t_1} = \mathcal{D}_x$.
Decomposing equation \eqref{1} grade by grade according to (\ref{g1}),  allows for obtaining the solution
for $D^{(i)} \in \lie_i$ (see for instance \cite{gui1}). This  is done
recursively,   starting from the highest grade projection of  \eqref{1}, 
namely 
\br
[E^{(1)}, \; D^{(N)} ]=0 \, ,
\label{highest}
\er
until   we arrive at the zero grade component,
\br
\pa_{t_{N}} A_0 - \pa_x D^{(0)} - [A_0, D^{(0)}]=0,
\label{2}
\er
which corresponds to the time evolution equation for the grade-zero field
$A_0$.

{Analogously} negative sub-hierarchy can be
constructed by considering negative graded decomposition : 
\br
[\mathcal{D}_x + A_x, \; \mathcal{D}_{t_{-N}}+ A_{t_{-N}}]= [ \mathcal{D}_x + E^{(1)} + A_0, \; \mathcal{D}_{t_{-N}}+D^{(-N )}  + D^{(-N +1)} \cdots + D^{(-1)} ]=0, \label{3}
\er
$ D^{(i)} \in \lie_i$.
 Different  gradings $Q$ and/or different $E^{(1)}$ lead to different 
integrable hierarchies.   The well-known examples
of AKNS and mKdV hierarchies,  
closely follow such framework for $\hat {\lie} = {\hat{sl}_2}$ 
algebra
but with  different   homogeneous and principal gradations, respectively
and consequently with different $E^{(1)}$ elements.

A universal  framework  unify positive and negative flows  can be formulated using the 
RHB factorization formula and we will see below that in more general
scheme presented in  \eqref{gr-h} it corresponds to
a conventional choice of $a=1$ .
In \cite{AGZ} this formula was introduced as: 
\br \label{rhb-conventional}
\exp \left( -\sum^{\infty}_{N=1} E^{(N)}t_{N}\right) \; 
g \; \exp \left( \sum^{\infty}_{N=1} E^{(-N)}t_{-N}\right)=
\Theta^{-1}_{-}\left(t \right)\Theta_{+} \left(t \right)\, .
\er
{In the above formula $E^{(\pm N)}  \in  \mathcal{K}^{(\pm N)} \equiv   
\mathcal{K} \cap \lie_{\pm N}$. For $N=1$ we encounter 
the grade one
semi-simple element $E^{(1)}$, a basic ingredient of a
zero-curvature approach. The group element
$g $  denotes an arbitrary constant group  element and 
$\Theta _- = \exp( -{\sum_{i>0}\theta^ {(-i)}}), \;\; 
\Theta_+ = \exp({\sum_{i\ge 0} \theta^ {(i)}})$.
In reference \cite{AGZ}  equation \eqref{rhb-conventional} was
used 
to derive the
Lax pair $A_x $ from a dressing factorization problem involving both
the positive and negative flows labeled by $\left(t_{\pm1},t_{\pm
2},\cdots \right)$ within the same formulation.  Both sub-hierarchies 
\eqref{1} and \eqref{3} were derived within such unified framework and the flows were shown to commute.}

\section{The Generalized Riemann-Hilbert-Birkhoff (g-RHB)  Decomposition}
\label{section:GRHB}

In
this paper we extend applicability of RHB
decomposition formula by generalizing the construction of \cite{AGZ} 
in a three separate ways: 
\begin{enumerate}
\item[(1)] First, we consider the construction of {\it  higher grading integrable hierarchies}  by 
replacing the grade $N$
semi-simple element $E^{(N)} \in \lie_N$ in equation \eqref{rhb-conventional}
by generalized  Heisenberg  generators (abelian) $\eps^ {(aN)}\in\lie_{aN}$.
\item[(2)]Apart  from  carrying higher graded structure  characterized by
the integer $a$, $\eps^{(aN)}$ may incorporate  information  about  non-zero vacuum induced by
 extra terms in $\eps^{(aN)}$ in addition to  $E^{(aN)}$ signaled  by non-zero value   of  parameter $b$  in equation (\ref{epsan}).
\item[(3)]
Incorporating explicitly a gauge ambiguity 
within the 
zero grade subspace that will be labeled by a parameter $c$ such that
$0 \le c \le 1$,  $ \tilde B  = e^ {-c\tilde {\theta}^{(0)} }$.
\end{enumerate}

The models, like KN, CLL, GI models
will emerge
when 
a higher grading semi-simple  element, e.g., $E = E^{(a)} \in \lie_a, 
\; a=2 $ is introduced. We refer to the  associated hierarchies as
{\it higher  grading  integrable hierarchies}.  
To incorporate these models in an algebraic approach  we propose
g-RHB decomposition formula that 
generalizes the RHB decomposition formula
as follows {\footnote{ In ref. \cite{AGZ} only negative/positive graded generators  were considered in the left/right hand sided of (\ref{rhb-conventional}). Here we consider both graded structure in each side of (\ref{rhb-conventional}) without loss of generality since  it  corresponds to redefining $\Theta_+$ and $\Theta_-$}} :
\br \label{gr-h}
  \Theta (t) =     \exp \left( -\sum^{\infty}_{N=1} \eps^{(-aN)}t_{-N}+\eps^{(aN)}t_{N}\right) \; 
    g \; \exp \left( \sum^{\infty}_{N=1}\eps^{(-aN)}t_{-N}+\eps^{(aN)}t_{N}\right)
    =\Theta^{-1}_{-} \left(t \right)\Theta_{+}  \left(t \right),\nonu \\ 
\er
where  $g $  again denotes an arbitrary group  element,  $a \in \mathbb{Z}_{+}^*$,      
$E^{(\pm aN)}  \in  \mathcal{K}^{(\pm aN)} \equiv   \mathcal{K} \cap \lie_{\pm aN}$, 
and the  constant elements $\eps^{(\pm aN)}$:
\begin{equation}
\eps^{(aN)} =   E^{(aN)}  + b \sum_{i=0}^{aN-1} \vareps_N^{(i)}, 
\label{epsan}
\end{equation}
\begin{equation}
\eps^{(-aN)} =   E^{(-aN)}  + b \sum_{i=1}^{aN-1} \vareps_{-N}^{(-i)}, \quad
\vareps_{\pm N}^{(i)} \in \lie_i,
\label{epsan2}
\end{equation}
describe  two possible vacuum structures
\begin{enumerate}
\item[I.] Type I hierarchies with {\it zero vacuum} solutions  and $\mathsf{b}=0$, $\eps^{(\pm a N)} =   E^{(\pm aN)} $.
\item[II.] Type II hierarchies  with {\it non-zero vacuum} solutions and $b=1$  in relations \eqref{epsan} { and \eqref{epsan2}.}
\end{enumerate}
In both cases the  elements $\eps^{(\pm a N)}$
satisfy  the center-less Heisenberg  (abelian ) 
subalgebra, 
\br 
[\eps^{(\pm aN )}, \, \eps^{(\pm a M)} ] =0,  \;\;\;\;\;\;\;\;\;  N, M \in \mathbb{Z}^*_+.\label{heisemberg}
\er 
 

     We now consider the g-RHB factorization (\ref{gr-h})  with  
   $\Theta(t)_-$ and  $\Theta_+(t)$  realized as : 
  \br
 \Theta_-(t)=\tilde B \Theta_<^{-1}=\tilde {B}\exp\left(-\sum_{k=1}^{\infty}
 \theta^{(-k)}\right) , \qquad \Theta_+(t)=\tilde {B}B\Theta_{>}=\tilde {B}B \exp\left(
 \sum_{k=1}^{\infty} \theta^{(k)}\right).  \label{theta-pi}
  \er
   for
 $ \tilde {B}$ being a zero grade element and independent of $B= \exp{ \theta_0}$.
 Also  $\theta^{(-k)} \in \lie_{-k}$ and
 $\theta^{(k)}\in \lie_{k}$, { where $k \in \mathbb{Z}_{+}$.} 

{It is important to notice the obvious fact that the g-RHB
decomposition formula (\ref{gr-h})} does not depend upon the  
$ \; \tilde B $. The zero grade  group element $\tilde B$ cancels out on
  the right hand side of equation (\ref{gr-h}) and we might as well gauge it away
or  alternatively set it to one : $\tilde B =1$. 
Its presence, however
turns to be important allowing definitions of different possible physical
 variables of the system.
We will  see below in
(\ref{ax}), (\ref{tn1}) that the addition of the
$\tilde B$  in (\ref{theta-pi}) reflects {a freedom to gauge
transform the model}  and therefore {exhibits an
ambiguity inherently} present in the formalism.


 In the remaining part of this section we will derive from a general decomposition
 of an affine Lie algebra the time evolution flow equations associated
 with algebra generators of positive and negative grade.


  The connection with vacuum parameters  of the theory
  given in  (\ref{1}) and (\ref{3}) in vacuum configuration, is established
  through:   
  \br
   & &  A_{t_{N}}^ {vac} = E^{(aN)} +  D^{(aN-1)}_{ vac} +\cdots D^ {(0)}_{ vac} \equiv   \eps^ {(aN)},  \;\; \;\; {\rm or } \label{x}\\
   & & A_{t_{-N}}^ {vac} = E^{(-aN)} +  D^{(-aN+1)}_{ vac}+ \cdots D^ {(-1)}_{ vac} \equiv \eps^ {(-aN)}. \label{N}
 \er
 $A_{x}^ {vac} =  A_{t_{1}}^ {vac}$.
Type I hierarchy is characterized by   {\it zero-vacuum } configuration
$ D^{(aN-1)}_{ vac} = \cdots D^ {(0)}_{ vac} =0$ or
$D^{(-aN+1)}_{ vac} = \cdots D^ {(-1)}_{ vac} =0 $
and  $ \eps^ {(\pm aN)} = E^ {(\pm aN)}$, i.e., $b=0$.
For {\it type II hierarchy}, the {\it  constant  non-zero vacuum}
configuration $ D_{vac}^{ (\pm aN \mp 1)}  \neq 0  $, i.e., $b=1$. In both cases   the zero curvature  representation  implies   $[ A_{x}^ {vac} ,A_{ t_{\pm N}}^ {vac} ]= [\eps^ {(a)}, \eps^ {(\pm aN)}]= 0$  and    from Jacobi identity we find that  (\ref{heisemberg}) follows  straightforwardly.
 Accordingly we conclude  that
 solution of  the zero curvature equation for the vacuum configuration
 can be  written , for both positive and negative sectors
as
\br A_x^{vac} = -\pa_x\Psi_0 \Psi_0^ {-1} = \eps^{(a)}, \qquad A_{t_{\pm N}}^{vac} = -\pa_{t_{\pm N}}\Psi_0 \Psi_0^ {-1}= \eps^ {(\pm aN)}, \label{avac}
\er
where  $\Psi_0 $ is the Baker-Akhiezer  function,
\br
\Psi_0 =  \exp \left( -\sum^{\infty}_{N=1} \eps^{(-aN)}t_{-N}+\eps^{(aN)}t_{N}\right), \quad x \equiv t_1.
\label{psizero}
\er

{In order to understand the commutativity of flows  and the   role of the parameter $a$ it
is instructive to define a submodel of the model defined by
equation \eqref{gr-h} with  $a=1$ by setting to zero the odd flows: }
\begin{equation*}
t_1=t_3={\ldots} = t_{2k+1}=0 ,
\end{equation*}
{on both sides of equation \eqref{gr-h}.
Then such
submodel of \eqref{gr-h} with $a=1$ agrees with the model obtained
by setting $a=2$ } {and $\eps^{(aN)}=E^{(aN)}$}
{in equation \eqref{gr-h} after an appropriate redefinition of
indices of the flows.
Similar relations hold for the negative times.
This, in principle, shows how to obtain the  $a=2,3, {\ldots} $
models as reductions of the
$a=1$ model and explains why their flows will commute.}

Acting with  derivatives  with respect to $t_1=x$ and $t_{\pm N}$ in 
equation (\ref{gr-h})  and using (\ref{avac}) we find, 
\begin{equation}
\pa_x \Theta = [\Theta, \; \eps^ {(a)}], \qquad \pa_{t_{\pm N}} \Theta =  [\Theta, \; \eps^ {(\pm aN)}],
  \label{dt}
\end{equation}
 which leads  from (\ref{x}) and (\ref{N}) to
 \br
 \Theta_{+} A_x^ {vac} \Theta_{+}^ {-1} -( \pa_x
 \Theta_{+})\Theta_{+}^ {-1} &=& \Theta_-A_x^ {vac}\Theta_-^ {-1} -
 (\pa_x \Theta_-)\Theta_-^ {-1}\,,     \label{aax1}\\
  \Theta_{+} A_{t_{\pm N}}^ {vac} \Theta_{+}^ {-1} -( \pa_{t_{\pm N}} \Theta_{+})\Theta_{+}^ {-1} &=& \Theta_-A_{t_{\pm N}}^ {vac}\Theta_-^ {-1} - (\pa_{t_{\pm N}} \Theta_-)\Theta_-^ {-1} . \label{atM}
 \er
 It therefore follows that the generators $\Theta_{\pm}$ can  be 
 identified with the dressing operators  mapping the vacuum into a non-trivial 
 configuration  by a gauge transformation, i.e., 
 \br
 A_x \equiv \Theta_{\pm} A_x^ {vac} \Theta_{\pm}^ {-1}- (\pa_x  \Theta_{\pm} ) \Theta_{\pm}^{-1}&=& -\pa_x (\Theta_{\pm}\Psi_0) \Psi_0^ {-1} \Theta_{\pm}^ {-1}, \label{aaxt} \\
 A_{t_{\pm N}} \equiv \Theta_{\pm} A_{t_{\pm N}}^ {vac} \Theta_{\pm}^ {-1}- (\pa_{t_{\pm N}}  \Theta_{\pm} ) \Theta_{\pm}^{-1}&=& -\pa_{t_{\pm N}} (\Theta_{\pm}\Psi_0) \Psi_0^ {-1} \Theta_{\pm}^ {-1},
\label{axt}
 \er
 It is clear from relation 
 (\ref{theta-pi}) and (\ref{gr-h})  that  the theory does not depend upon 
 $\tilde B$. 
 This means that {\it different choices of  $\tilde B$  leads to the same hierarchy}. This  fact is expressed  at the zero curvature level from (\ref{aaxt}) and  (\ref{axt}) by gauge invariance under transformation
  $\Theta_{\pm} \rightarrow \tilde B \Theta_{\pm}$, i.e.,
 \br
  \tilde A_x& =&\tilde B A_x \tilde B^{-1} - \pa_x \tilde B \tilde B^ {-1}, \label{gx} \\
  \tilde A_{t_{\pm N}}& =&\tilde B A_{t_{\pm N}} \tilde B^{-1} - \pa_{t_{\pm N}} \tilde B \tilde B^ {-1}. \label{gt}
\er

 From (\ref{aax1}) and (\ref{atM})  we find ( recall $N\in Z^{*}_+$),
 \br 
 -\pa_{t_{N}}\Theta_- \Theta_{-}^{-1} + \pa_{t_N} \Theta_+
 \Theta_{+}^{-1} &=& \Theta_{+} \eps^ {(aN)} \Theta_+ - \Theta_- \eps^
 {(aN)}\Theta_-^{-1}\, , \label{-}  \\
  -\pa_{t_{-N}}\Theta_- \Theta_{-}^{-1} + \pa_{t_{-N}} \Theta_+
  \Theta_{+}^{-1} &=& \Theta_{+} \eps^ {(-aN)} \Theta_+ - \Theta_-
  \eps^ {(-aN)}\Theta_-^{-1}\, . \label{+}
 \er
 Considering the graded structure of $\Theta_{\pm}$ in
 (\ref{theta-pi}) we find by projecting (\ref{-}) and (\ref{+}) into zero grade:
\br 
  (\pa_{t_N}\tilde B )\tilde B^ {-1} - (\pa_{t_N}\tilde B  B )B^ {-1} \tilde B^ {-1}= -\tilde B(\pa_{t_N} B ) B^ {-1} \tilde B^ {-1}&=&\( \Theta_- \eps^ {(aN)} \Theta_-^ {-1}\)_0, \label{310}\\
 (\pa_{t_{-N}}\tilde B )\tilde B^ {-1} - (\pa_{t_{-N}}\tilde B  B )B^
 {-1} \tilde B^ {-1}= -\tilde B(\pa_{t_{-N}} B ) B^ {-1}\tilde B^ {-1} &=&-\( \Theta_+ \eps^ {(-aN)} \Theta_+^ {-1}\)_0\,. \label{311}
 \er
  Acting with derivatives in $\Theta_{\pm}$  we obtain,
 \br
 (\pa_{t_{\pm N}}\Theta_- )\Theta_-^ {-1} &=& \pa_{t_{\pm N}}( \tilde
 B \Theta_<^{-1})\Theta_-^ {-1} \nonu \\
 &=&  (\pa_{t_{\pm N}}\tilde B) \tilde
 B^{-1} + \tilde B  (\pa_{t_{\pm N}}\Theta_<^{-1}) \Theta_< \tilde B^
 {-1} \, ,\label{01}\\
 (\pa_{t_{\pm N}}\Theta_+ )\Theta_+^ {-1} &=& \pa_{t_{\pm N}} (\tilde B B \Theta_>)\Theta_+^ {-1} \nonu \\
 &=&(\pa_{t_{\pm N}}\tilde B) \tilde B^{-1} +\tilde B (\pa_{t_{\pm N}} B)  B^{-1}\tilde B^ {-1} +
  \tilde B  B(\pa_{t_{\pm N}}\Theta_>) \Theta_>^ {-1} B^ {-1}\tilde B^
  {-1} \, .\nonu \\
  \label{02} 
 \er
From relations (\ref{-}) ,  (\ref{+}),   (\ref{01})  and   (\ref{02}) we find
 \br
 \pa_{t_N}B&=&B \(\Theta_>\eps^{(aN)}\Theta^ {-1}_>\)_0 -
 \(\Theta_<^{-1}\eps^{(aN)}\Theta_<\)_0 B \, , \label{b1}\\
 \pa_{t_{-N}}B&=&B\(\Theta_>\eps^{(-aN)}\Theta^ {-1}_>\)_0\, ,\label{b2} \\
 \pa_{t_N}\Theta_>&=&\(\Theta_>\eps^{(aN)}\Theta^ {-1}_>\)_> \Theta_>-
 B^ {-1} \(\Theta_<^ {-1}\eps^{(aN)}\Theta_<\)_> B \Theta_> \, ,\\
 \pa_{t_{-N}}\Theta_>&=&\(\Theta_>\eps^{(-aN)}\Theta_>^{-1}\)_>\Theta_>
 \, ,\\
\pa_{t_N}\Theta_<&=&-\Theta_<\(\Theta_<^ {-1}\eps^{(aN)}\Theta_<\)_<
\, ,\\
 \pa_{t_{-N}}\Theta_<&=& \Theta_> B\(\Theta_>\eps^{(-aN)}\Theta_>^
 {-1}\)_< B^{-1}\Theta_<- \Theta_<\(\Theta_<^
 {-1}\eps^{(-aN)}\Theta_<\)_<   \, .
\er
 As a consequence, from equations (\ref{01}) and (\ref{02})  we find 
  \br 
  \pa_{t_N}\Theta_+&=&\(\pa_{t_N}\tilde B \tilde B^ {-1}+\Theta_+\eps^
  {(aN)} \Theta_+^ {-1} - (\Theta_- \eps^ {(aN)} \Theta_-^
  {-1})_{\ge}\)\Theta_+  \, ,\label{a1}\\
  \pa_{t_{-N}}\Theta_+ &=&\( \pa_{t_{-N}}\tilde B \tilde B^ {-1} +
  \Theta_+\eps^ {(-aN)}\Theta_+^ {-1} - ( \Theta_+\eps^
  {(-aN)}\Theta_+^ {-1} ) _<  \) \Theta_+  \, , \label{a2}\\
   \pa_{t_N}\Theta_-&=& \(\pa_{t_N}\tilde B \tilde B^
   {-1}+\Theta_-\eps^ {(aN)} \Theta_-^ {-1} - (\Theta_- \eps^ {(aN)}
   \Theta_-^ {-1})_{\ge}\)\Theta_-  \, ,\label{a3}\\
  \pa_{t_{-N}}\Theta_- &=&\( \pa_{t_{-N}}\tilde B \tilde B^ {-1} +
  \Theta_-\eps^ {(-aN)}\Theta_-^ {-1} - ( \Theta_+\eps^
  {(-aN)}\Theta_+^ {-1} ) _<  \) \Theta_-  \, .\label{a4}
   \er
 Introducing, $\mathcal{D}_{t_N} \Theta_- = \pa_{t_N} \Theta_- + \Theta_- \mathcal{D}_{t_N}$, we find from (\ref{a3}),
  \br
 \Theta_- (\mathcal{D}_{t_N} + \eps^ {(aN)} )\Theta_-^{-1} =
 \mathcal{D}_{t_N} + (\Theta_-\eps^ {(aN)}\Theta_-^{-1})_{\ge} -
 (\pa_{t_N} \tilde B)\tilde B^ {-1} \, .
 \er

 From the abelian nature of $\cal{ K_{\eps}} $, i.e., $[\eps^ {(a)}, \; \eps^ {(aN)} ] = 0$ and  denoting $t_1=x$,  we find
 \br
 \Theta_- [ \mathcal{D}_x + \eps^{(a)}, \; \mathcal{D}_{t_N} + \eps^
 {(aN)}]\Theta_-^{-1} = [ \Theta_-(\mathcal{D}_x
 +\eps^{(a)})\Theta_-^{-1},\;  \Theta_-(\mathcal{D}_{t_N}
 +\eps^{(aN)})\Theta_-^{-1}]=0 \, .
 \er
 
It therefore follows that
\br
[ \mathcal{D}_x + \tilde A_x, \; \mathcal{D}_{t_{N}} + \tilde A_{t_N}
]=0 \, ,
\er
where
\br 
\tilde A_x &=& \tilde B (\Theta_<^ {-1} \eps^{(a)} \Theta_<)_{\ge}
\tilde B^{-1}- (\pa_x \tilde B) \tilde B^{-1} \equiv \tilde B
E^{(a)}\tilde B^{-1}+\sum_{i=0}^{a-1} \tilde A_i \, ,\label{ax} \\ 
 \tilde {A}_{t_N} &=& \tilde B (\Theta_<^ {-1} \eps^{(aN)}\Theta_<)_{\ge} 
 \tilde B^{-1} - (\pa_{t_N} \tilde B) \tilde B^{-1} \equiv
 \sum_{i=0}^{aN} \tilde D^ {(i)}\, . \label{tn1}
 \er
 It thus follows from (\ref{gx}) and (\ref{gt})  that 
 \br
 A_x &=& (\Theta_<^ {-1} \eps^{(a)} \Theta_<)_{\ge}= E^{(a)}+ A_{a-1} + \cdots A_0, \label{axx} \\
 {A}_{t_N} &=&  (\Theta_<^ {-1} \eps^{(aN)}\Theta_<)_{\ge}=\ D^ {(aN)}
 + D^ {(aN-1)} + \cdots D^ {(0)} \, . \label{tnn}
 \er
For the negative graded sector we find,
\br
 \Theta_- [ \mathcal{D}_x + \eps^{(a)}, \; \mathcal{D}_{t_{-N}} +
 \eps^ {(-aN)}]\Theta_-^{-1} = [ \Theta_-(\mathcal{D}_x
 +\eps^{(a)})\Theta_-^{-1},\;  \Theta_-(\mathcal{D}_{t_{-N}}
 +\eps^{(-aN)})\Theta_-^{-1}]=0 \, ,
 \er
 leading to
\br
[ \mathcal{D}_x + \tilde A_x, \; \mathcal{D}_{t_{-N}} + \tilde
A_{t_{-N}} ]=0 \, , 
\er
where
\br
\tilde {A}_{t_{-N}} &=& \tilde B
B(\Theta_>\eps^{(-aN)}\Theta_>^{-1})_{<} B^{-1} \tilde B^{-1} -
(\pa_{t_{-N}} \tilde B) \tilde B^{-1}  \equiv \sum_{i=0}^{aN} \tilde
D^ {(-i)}  \, . \label{mtn}
 \er
For the quantity  $A_{t_{-N}}$   from (\ref{gt})  we find ,
\br 
 {A}_{t_{-N}} = B (\Theta_>\eps^{(-aN)}\Theta_>^{-1})_{<} B^{-1}  = D^{(-aN)} + D^ {(-aN+1)} \cdots D^ {(-1)}.  \label{atneg}
\er
{ Notice that  (\ref{atneg}) does not contain  zero-grade terms as in
(\ref{tnn}) due to the asymmetric splitting  of zero grade subgroup $B$ in 
(\ref{theta-pi}) when $\tilde B=1$.
The absence of  such terms  will be responsible for excluding the 
possibility of constant vacuum solutions  for certain flows.
This  will explain why  flows   associated  to $t_{-N}=t_{-1}$  only   allows  zero vacuum solutions in the  explicit examples   to be discussed  in the next sections. }
We should also point out  that the same result can be obtained using 
$\Theta_+$  from (\ref{a1}) and ( \ref{a2}) instead of $\Theta_-$.  The  group element $\tilde B$ act as a gauge freedom which can be chosen  by convenience. In contrast,  the zero grade group element $B$ contains physical fields and satisfy (\ref{b1}) and (\ref{b2}).  In particular  for $N=1$,
\br 
\pa_x B B^ {-1} =b B \vareps^{(0)}_1 B^{-1}- (\Theta_<^ {-1} \eps ^{(a)} \Theta_<)_0. \label{cd}
\er



\section{Integrable Hierarchies obtained from g-RHB decomposition}
\label{section:models}
Here we will illustrate our construction
for the case of the affine algebra  $\hat{\lie} =\hat {sl}_2$.
Neglecting for the moment  the central terms, the  nonzero  commutators are {\footnote{ In centerless case  (loop algebra) the affine generators are realized as $h^ {(n)} = \zeta^ n h, E_{\pm \a}^ {(n)} = \zeta^ n E_{\pm \a}$ and $d = \zeta \frac{ \pa}{\pa\zeta}$.}}
\br \[ h^{(n)}, \,  E_{\pm \a}^{(m)}\] = \pm 2  E_{\pm \a}^{(n+m)}, \,\, 
\[ E_{\a}^{(n)}, \, E_{- \a}^{(m)}\] =h^{(n+m)},\,\,
\[ d ,\, h^{(n)} \] = n h^{(n)},\,\, \[ d ,\; E_{\pm\a}^{(n)} \]= n E_{\pm\a}^{(n)},
\er
{ where $n,m \in \mathbb{Z}$.}

The structure of the two dimensional gauge potentials (\ref{ax}),
(\ref{tn1}) and (\ref{mtn}) naturally decomposes into two
sub-hierarchies namely, {\it positive and negative} according to the
different gradings as we shall now discuss explicitly  case by case.
 
We will work here exclusively with the  principal gradation, with
$Q = Q_{p} = 2
 d + \frac{1}{2}h$ {such that}     $ [Q_p, \lie_i ] = i \lie_i$   where,
  \br
   \lie_{2n} = \{ h^ {(n)}
 \}, \;\;\qquad \lie_{2n+1} = \{ E_a^{(n)}, E_{-\a}^ {(n+1)} \} \, ,\label{ppal}
 \er 
 and $ \hat{\lie} = \oplus_{i \in \mathbb{Z} } \lie_i $.

 \subsection{The  $a=1$ mKdV Hierarchy  }
 \label{subsection:mkdv}
For completeness we  reproduce here the well-known mKdV 
hierarchy  in the setting considered in this paper.
Set $ a=1$ and $E \equiv E^{(1)}  = \frac12(E_{\a}^{(0)} + E_{-\a}^{(1)})$. 
 It follows that the kernel $\cal {K}$ of $E$  contains only  {\it odd graded} generators 
 and is given by
 \br
 Ker_{E} = { \cal {K}}= \left\{  E^{(2n+1)} = \frac12\(E_{\a}^{(n)} + E_{-\a}^{(n+1)}\) \in \lie_{ 2n+1} \right\}
 \er
   (see \cite{gui1} for positive and negative flows).

   In fact  $E^{(1)} $ is a semi-simple element in the sense that 
   ${\hat {\lie }}= \cal {K} \oplus {\cal {M}} $ where ${\cal {M}}$ is
   an image of $E$.
 In this case $E^{(\pm N)} $ satisfy the centerless Heisenberg algebra,
 \br
 [ E^{(2n+1)}, \; E^{(2m+1)}]=0.  \label{eps-hei}
 \er

An important  second Heisenberg algebra satisfying (\ref{eps-hei}) is generated by the combination {\footnote{ Notice that $\Sigma$ contain terms of different gradings.
 It does suggest that powers of $v_0$ could be incorporated into a generalized grading operator such that the two terms  would have the same grade. 
 The vacuum parameter $v_0$ would introduce an  extra loop in the affine algebra (see for instance ref. \cite{twoloop}). }}
\br 
\Sigma^ {(2n+1)} = \frac{1}{2} (E_\a^ {(n)} + E_{-\a}^ {(n+1)}) + bv_0
h^ {(n)} \,.\label{rho}
\er
Denoting  $\Sigma^ {(1)} =\eps^{(1)} = E + bv_0 h^{(0)},\;  \; \theta^{(-1)} =
 \chi_{-1} E_{\a}^{(-1)} + \psi_{-1}E_{-\a}^{(0)}$ and 
 $\tilde  B = e^{-c\tilde \phi h^{(0)}}$, we obtain from
 (\ref{ax}): 
 \br 
 \tilde A_x&=&\tilde B \(E^{(1)}+ b v_0 h^{(0)}- [ {\theta^{(-1)}},
 {E^{(1)}} ] \) \tilde B^{-1}-\tilde B \pa_x \tilde B^{-1} \nonu \\  &=& \frac12 \(e^{-2c\tilde \phi(t)}
 E_{\alpha}^{(0)}+e^{2c\tilde \phi(t)}E_{-\alpha}^{(1)}\)+ \frac12 \( -\chi_{-1}+\psi_{-1}  
 +2b v_0+2c\pa_x \tilde \phi \) h^{(0)} \, .
 \label{ax1}
 \er 
We set $A_0 = \frac{1}{2}( -\chi_{-1}+\psi_{-1}  
 +2b v_0)h^{(0)}= v(x,t_{\pm N})h^{(0)}$ and notice
that $\tilde B A_0 \tilde B^{-1} =A_0$ and   henceforth the 
field $v(x,t_{\pm N})$ is  {\it unchanged under gauge transformation}  (\ref{gx}) 
and (\ref{gt}).
As a consequence, although  the gauge potentials $A_x$ and $A_{t_{\pm N}}$ may 
depend upon $\tilde B$,  the flow equations  are independent of $\tilde B$ and we may take 
$\tilde B=1, \; \;$ or $c=0$, to  obtain:
\br  A_x = E + A_0, \qquad A_0 = \frac{1}{2}( -\chi_{-1}+\psi_{-1}  
 +2bv_0)h^{(0)}= v(x,t_{\pm N})h^{(0)}. \label{mkdvlax}
\er
 Equation (\ref{mkdvlax}) shows that the  
  extra term in $\eps^{(1)}$, namely $ \vareps_1^{(0)} =  v_0 h^{(0)}$   in (\ref{epsan}) 
  represents the non-zero vacuum  value $v_0$, for $b\neq 0$ 
  characteristic for the type II hierarchy. 
 
From the general structure of $A_{t_N}$ from (\ref{tn1})  we find the zero curvature,
\br
   \[ {\mathcal{D}_{x}+E^{(1)}+A_{0}},\; {\mathcal{D}_{t_{N}}+D^{(N)}+\sum^{N-1}_{i=0}D^{(i)}}\]
   =0 \, ,\label{zcc1}
\er
which decomposes   grade by grade  allowing to determine 
$D^{(i)} = D^{(i)}_{{\cal K}} + D^{(i)}_{{\cal M}}$.
In particular the highest grade   projection 
\begin{equation*}
   \[ {E^{(1)}}, \; {D^{(N)}}\] =0,
\end{equation*}
leads to ${D^{(N)}} = D^ {(N)}_{{\cal {K}}} \in \mathcal{K}$ and
hence  for the positive sub-hierarchy it holds that $ N=2k+1 $.  The second  highest grade projection,
\begin{equation*}
[E^{(1)}, D^{(N-1)} ] + [A_0, \; E^{(N)} ] + \pa_x D^{(N)} =0
\end{equation*}
determines $ D^{(N-1)}_{{\cal M}}\in {\cal M}, \cdots $. This process continues
until we reach the zero grade  projection,
\begin{equation*}
\pa_x D^{(0)} -\pa_{t_N} A_0 + [A_0,\; D^{(0)} ] =0 \, ,
\end{equation*}
which  yields the flow equation  for $v(x, t_N)$.

Solving (\ref{zcc1}) for $t_{3}$ and $t_5$ we find,
\begin{align}
A_{t_3}&=\frac{1}{2} \left (E_{\alpha}^{1} +  E_{-\alpha}^{2}\right )+v h^{(1)}-\left (2v^3-v_{xx} \right )h^{(0)}  \nonu \\
    &- \left (v^2-v_{x}  \right ) E_{\alpha}^{(0)}
    - \left (v^2+v_{x}  \right )E_{-\alpha}^{(1)} \label{a33}
\end{align}
and 
\br A_{t_5}&=& \, \frac{1}{2} \left( E_{\alpha}^{(2)} +   E_{-\alpha}^{(3)}\right) +v h^{(2)} +\left( v_{x}-v^2\right) E_{\alpha}^{(1)} - \left( v_{x}+v^2\right)  E_{-\alpha}^{(2)} \nonu \\
    &+&\left( v_{xx}- 2 v^3\right) h^{(1)}+\left [ v_{3x}-2vv_{xx}+v_x\left ( v_x-6v^2 \right ) +3v^4 \right ] E_{\alpha}^{(0)} \nonu  \\
    &-&\left [ v_{3x}+2vv_{xx}-v_x\left ( v_x+6v^2 \right ) -3v^4 \right ]  E_{-\alpha}^{(1)} \nonu \\
    &+&\left [ v_{4x}-2v\left (5vv_{xx}+ 5v_{x}^2-3v^4 \right ) \right ] h^{(0)} \label{a55}
\er
and arrive at the $t_3$  (mKdV ) and $t_5$ flows:  
\begin{equation}
\begin{aligned}
    \partial_{t_3}v=\partial^3_x v-6v^2 \pa_x v \, ,
    \label{mkdv}
\end{aligned}
\end{equation}
and 
\begin{equation*}
\pa_{t_5}v =\pa_x^5 v - 10 v^2 \pa_x^3v-40v \pa_x v \pa_x^2v -10 (\pa_xv)^3 +30 v^4 \pa_xv. \label{t5}
\end{equation*}
Higher projections of  the zero curvature equations \eqref{zcc1}  lead to  higher flow equations  for $t_N$ .  
It is clear  that the {\it positive odd} sub-hierarchy admits
{\it  both, zero $v=0$  and constant non-zero vacuum, $v=v_0 \neq 0$}  structures.  In vacuum configuration  we find for $t_3$ and $t_5$, 
\br
A^ {vac}_{t_3}&=& \frac{1}{2}(E_{\alpha}^{(1)} +  E_{-\alpha}^{(2)}) +b v_0h^ {(1)} - 2bv_0^ 2 \( \frac{1}{2}(E_{\alpha}^{(0)} +  E_{-\alpha}^{(1)}) + bv_0h^{(0)}\) \nonu \\
&=& \Sigma^ {(3)} - 2bv_0^ 2 \Sigma^{(1)}\, ,\label{t3vac}
\er
\br
A^ {vac}_{t_5}&=& \frac{1}{2}(E_{\alpha}^{(2)} +  E_{-\alpha}^{(3)}) +bv_0h^ {(2)} - 2bv_0^ 2 \( \frac{1}{2}(E_{\alpha}^{(1)} +  E_{-\alpha}^{(2)}) + bv_0h^{(1)}\) \nonu \\
&+& bv_0^ 4 \( \frac{1}{2}(E_{\alpha}^{(0)} +  E_{-\alpha}^{(1)}) + b v_0h^{(0)}\) \nonu \\
&=& \Sigma^ {(5)} - 2bv_0^ 2 \Sigma^{(3)}+6 bv_0^ 4 \Sigma^ {(1)} \, .\label{t5vac}
\er
In general  we expect a the following structure 
\br 
A^ {vac}_{t_{2k+1}} = \sum _{j=0}^{k} c_j (bv_0)^{2k-2j}\Sigma^ {(2j+1)}, \quad k = 0, 1, 2,  \cdots\label{avacpos}
\er
  for the positive sub-hierarchy, where $c_j$ are constants independent of $v_0$. It is interesting to observe that 
  the sum the powers of $v_0$ and  the grade of $\Sigma$   always add up to  $2k+1$.

The very same argument  works for negative flows, i.e.  we  can solve  the zero curvature  condition  and determine  a sub-class  of negative hierarchy   for $b =0$. From (\ref{mtn}),
\br
   \[ {\mathcal{D}_{x}+E^{(1)}+A_{0}},\; {\mathcal{D}_{t_{-2k-1}}+B E^{(-2k-1)} B^{-1} 
      +\sum^{2k}_{i=1}D^{(-i)}}\]=0.\label{zccneg1}
\er
The simplest case  of  $k=0$  and $b=0$ has,
\br
A_{t_{-1}}= \frac{1}{2}\left( e^{-2 \pa^{-1}_x v} E_{\a}^{(-1)} + e^{2 \pa^{-1}_x v} E_{-\a}^{(0)} \right), \label{atm1}
\er 
leading to the sinh-Gordon equation i.e., \footnote{Defining $B=e^{\phi(t) h^{(0)}}$ with $\phi(t)=\phi'(t)-bv_0x$, the equation \eqref{cd} implies $\pa_x\phi=-v$.}
\br
\pa_{t_{-1}}v = \frac{1}{4}\left(e^{2 \pa^{-1}_xv} -e^{-2 \pa^{-1}_xv}\right) \;  \longrightarrow \;\;  \pa_{t_{-1}}\pa_x \phi  = \frac{1}{4}\left(e^{2 \phi } -e^{-2 \phi }\right), \quad -\pa_x \phi  = v
\label{tm1}
\er
Next for $t_{{-3}}$ we find,
\br
\pa_{t_{-3}}\pa_x \phi = \frac{1}{8}e^{2\phi} \pa_x ^{-1}\( e^{-2\phi} \pa_x^{-1} (e^{2\phi}-e^{-2\phi})\) + \frac{1}{8}e^{-2\phi} \pa_x ^{-1}\( e^{2\phi} \pa_x^{-1} (e^{2\phi}-e^{-2\phi})\).
\label{tm3}
\er
All the  {\it odd  graded } negative  flows,   are constructed from {the} {\it zero vacuum configuration, } of
{ $b = 0, \; v=0, \; \phi=0$} and the Lax
potentials
 \br
 A^{vac}_x = E^{(1)} 
 \er
  and 
\br
A^{vac}_ {t_{-2k-1}}  =  E^{(-2k-1)} =\frac{1}{2} (E_{\a}^ {(-k-1)}+ E_{-\a}^ {(-k)}), \quad k = 0, 1, 2,  \cdots
\er
satisfying the Heisenberg subalgebra (\ref{eps-hei}) 
as in (\ref{gr-h}) with $a=1$ and identified with { \it Type I  hierarchy. }

Recall that as  it was pointed out in \cite{gui1}   
for the negative flows there  is no restriction   upon the values of  $N$.
There exists  a 
second sub-hierarchy  consisting
of {\it  negative even} grades (see   \cite{gui1}, \cite{qiao}).
{ In this case  we can find a  second Heisenberg subalgebra  by
considering non-zero constant vacuum solutions e.g., $A_x^{vac}= \Sigma^{(1)}\equiv
E^{(1)} + v_0 h^{(0)}$.
In fact such  class of integrable models  admits strictly  {\it non zero constant vacuum} 
solutions $ v=v_0 \neq 0,$ and is referred to as {\it Type II hierarchy}. 
Let us consider, apart from the positive odd flows, the  negative even 
flows. The potential :
\br 
A_{t_{-2}} = \frac{1}{2} h^{(-1)} + \frac{1}{2} e^{2\phi}\pa_x^{-1} e^{-2\phi}\; E_{\a}^{(-1)} - \frac{1}{2} e^{-2\phi}\pa_x^{-1} e^{2\phi} \; E_{-\a}^{(0)}.
\er 
leads to the $t_{-2}$ flow,
\br 
\pa_{t_{-2}} \pa_x \phi =\frac{1}{4}e^{2\phi} \pa_x ^{-1} e^{-2\phi} 
+\frac{1}{4}e^{-2\phi} \pa_x ^{-1} e^{2\phi} .\label{tm2}
\er
Similarly we find  the $t_{-4}$ flow, 
\br
\pa_{t_{-4}} \pa_x \phi =\frac{1}{8}e^{2\phi} \pa_x ^{-1} \left[e^{-2\phi} \left(\partial_x^{-1}e^{2\phi} \right)\left(\partial_x^{-1}e^{-2\phi} \right)  \right]+\frac{1}{8} e^{-2\phi }\pa_x ^{-1} \left[e^{2\phi}
\left(\partial_x^{-1}e^{2\phi} \right)\left(\partial_x^{-1}e^{-2\phi} \right)  \right] .\nonu \\\label{tm4}
\er
It is straightforward to verify that $\phi= -v_0x$  indeed
satisfies equations (\ref{tm2}) and (\ref{tm4}) for $v_0\neq
0$. The vacuum structure of such a class of models implies
a $v_0$-deformation upon the algebraic structure.
The vacuum structure is given by
\br
A_{t_{-2}}^{vac} 
&=& \frac{1}{2 v_0} \( \frac{1}{2 }(E_\a^ {(-1)}+ E_{-\a}^ {(0)})
+ v_0 h^{(-1)}\)=  \frac{1}{2 {v_0}}\Sigma^{(-1)}\, ,\nonu \\
A_{t_{-4}}^{vac} 
&=& \frac{1}{2 v_0} \( \frac{1}{2 }(E_\a^ {(-2)}+ E_{-\a}^ {(-1)})
+ v_0 h^{(-2)}\)=  \frac{1}{2 {v_0}}\Sigma^{(-3)},\nonu \\
&\vdots & \nonu \\
A_{t_{-2k-2}}^{vac} 
&=& \frac{1}{2 v_0} \( \frac{1}{2 }(E_\a^ {(-k-1)}+ E_{-\a}^ {(-k)})
+ v_0 h^{(-k-1)}\)=  \frac{1}{2 {v_0}}\Sigma^{(-2k-1)}, \quad k = 0, 1, \cdots \label{avacneg}
\er

One and two soliton solutions  were constructed for the entire hierarchy  in terms of    deformed vertex operators in \cite{gui1}.  
These  are  $v_0$ dependent eigenfunctions of  $A_{t_{\pm N}}^{vac}\quad N= 1, \cdots $  and the  soliton dispersion relations are given by their $v_0$ dependent eigenvalues.



 
\subsection {Higher Grading $a=2$  Hierarchies -  KN, GI and  CLL hierarchies. } 
\label{subsection:kn}

Let us now set   $ a=2 $ and, as a  first example consider the GI
hierarchy generated from the semisimple element:  
 \br
\eps^{(2)} =  \frac{1}{2} h^{(1)} + b r_0 E_{\a}^{(0)}+b s_0 E_{-\a}^{(1)} -br_0 s_0 h^{(0)}.
 \label{gi-eps}
\er


 From equation (\ref{ax}) we find
that
$ A_x =
  (\Theta_<^ {-1} \eps^{(2)} \Theta_<)_{\ge}$ is given by 
\br
\label{axgi}	
    A_x = \eps^{(2)} - [ \theta^{(-1)}, E^{(2)}]-b [ \theta^{(-1)},  r_0 E_{\a}^{(0)}+ s_0 E_{-\a}^{(1)}] + \frac{1}{2} \[ \theta^{(-1)}, 
    \[ \theta^{(-1)}, E^{(2)} \]\] \, ,
\er
where $\theta^{(-1)} =
 \chi_{-1} E_{\a}^{(-1)} + \psi_{-1}E_{-\a}^{(0)}, \quad \theta^{(-2)} =
 \phi_{-2}h^{(-1)} $.
 We therefore find 
 \br
 A_x=\frac{1}{2}h^{(1)}
 +r(t)E_{\alpha}^{(0)}+s(t) E_{-\alpha}^{(1)} - r(t)s(t) h^{(0)},\label{axrs}
 \er
 where  we have defined the field variables ($b^2=b$)
\br
r(x, t_{\pm N})=br_0+\chi_{-1}(x, t_{\pm N}), \qquad s(x,t_{\pm N})=bs_0 -\psi_{-1}(x, t_{\pm N}).
\er
 The flow equations are  derived  by solving the zero curvature relation:
  \begin{equation*}
   \[ \mathcal{D}_{x}+\frac{1}{2}h^{(1)}+r \left(t\right)
   E_{\alpha}^{(0)}+s\left(t\right)E_{-\alpha}^{(1)} -
   r \left( t \right)s \left(t\right)h^{(0)}, \;\; 
   \mathcal{D}_{t_{N}}+E^{(2N)}+\sum^{2N-1}_{i=0}D^{(i)} \] =0.
\end{equation*}
 
 Solving for $N=2$ we find { \footnote{where the constant term is fixed by a choice of integration constants.}}
 \br
  A_{t_2}=\frac{1}{2} h^{(2)}  +r E_{\alpha}^{(1)} +  s E_{-\alpha}^{(2)}  -r s h^{(1)}  -r_x E^ {(0)}_{\a} + s_x E^ {(1)}_{-\a}
  -\( r^2s^2-s r_x +r s_x -br_0^2 s_0^2\) h^{(0)}.
        \er
        Analogously for $N=3$,
\br
    A_{t_3}&=&\frac{1}{2} h^{(3)}  +r E_{\alpha}^{(2)} +  s E_{-\alpha}^{(3)}
  -r s h^{(2)}  -r_x E^ {(1)}_{\a} + s_x E^ {(2)}_{-\a}
  -(r s_x- sr_x+r^2s^2) h^{(1)}\nonu \\
  &+&\( r_{xx}-2r^2(s_x+rs^2) \)E_\a^{(0)}
  +\( s_{xx}+2s^2(r_x-r^2s) \) E_{-\a}^{(1)}\nonu \\
  &-&(rs_{xx}+r_{xx}s-r_xs_x-2r^3s^3)h^{(0)}\, ,\nonu \\
\er
 leading to 
  \br
     \partial_{t_2}r=-\partial^2_xr+2  
r(r^2s^2-br_0^2 s_0^2) +2
r^2s_x, \qquad
   \partial_{t_2}s =\partial^2_xs-2 s(r^2s^2-br_0^2 s_0^2)+2s^2r_x \, , \label{gi2}
\er
and
\br
\pa_{{t_3}} r=\partial^3_xr-6rr_x(rs^2+s_x), \qquad
\pa_{{t_3}}s=\partial^3_xs-6ss_x(r^2s-r_x) \, .\label{gi3}
\er
Other flow equations can be systematically derived from
(\ref{tn1}) for higher values of $N$ and together they form the
GI positive hierarchy. Equations (\ref{gi2}) admit  
{\it zero vacuum} solution for $b=0$ and  {\it constant vacuum} for
$b =1$. Notice that explicit vacuum parameters appear explicitly in
eq. (\ref{gi2}) due to the fitting with of the $b$ dependent terms from the
Heisenberg operators $\eps^{(aN)}(b)$.  Other equations, like
(\ref{gi3}) do not display explicit $b$ dependent terms and admit
both {\it zero} and {\it constant non-zero } vacuum solutions.

For the negative GI sub-hierarchy   we solve the zero curvature equation
 \begin{equation*}
   \[ \mathcal{D}_{x}+\frac{1}{2}h^{(1)}+r \left(t\right)
   E_{\alpha}^{(0)}+s\left(t\right)E_{-\alpha}^{(1)} -
   r \left( t \right)s \left(t\right)h^{(0)}, \;\;
   \mathcal{D}_{t_{-N}}+\sum^{2N}_{i=1}D^{(-i)} \] =0.
\end{equation*}

For the first few  cases we find,
\br
A_{t_{-1}}&=& \frac{1}{2}h^{(-1)}+e^{2\phi} \pa_x^{-1}\(re^{-2\phi}\) E_\a^{(-1)}-e^{-2\phi} \pa_x^{-1}\(se^{2\phi}\) E_{-\a}^{(0)}, \label{atm1gi}
\er
and
\br
A_{t_{-2}}&=& \frac{1}{2}h^{(-2)}+e^{2\phi} \pa_x^{-1}\(re^{-2\phi}\) E_\a^{(-2)} -e^{-2\phi} \pa_x^{-1}\(se^{2\phi}\) E_{-\a}^{(-1)} \nonu \\
&+& \pa_x^{-1}\(re^{-2\phi}\) \pa_x^{-1}\(se^{2\phi}\) h^{(-1)} \nonu \\
&-&e^{2\phi} \partial_x^{-1}\left[\partial_x^{-1}\left( r e^{-2\phi}\right)-2re^{-2\phi}\partial_x^{-1}\left( r e^{-2\phi}\right) \partial_x^{-1}\left( s e^{2\phi}\right) \right]E_{\a}^{(-1)} \nonu \\
&-&e^{-2\phi}\partial_x^{-1}\left[\partial_x^{-1}\left( s e^{2\phi}\right)+2se^{2\phi}\partial_x^{-1}\left( r e^{-2\phi}\right) \partial_x^{-1}\left( s e^{2\phi}\right) \right]E_{-\a}^{(0)},
\label{atm2gi}
\er
where $\pa_x \phi=rs$ \footnote{Defining $B=e^{\phi(t) h^{(0)}}$ with $\phi(t)=\phi'(t)+br_0s_0x$ , the equation \eqref{cd} implies $\pa_x\phi=rs$.}, leading to the following flow equations,
\br
\pa_{t_{-1}}r = e^{2\phi}\pa_x^{-1}\(r e^{-2\phi}\), \qquad
\pa_{t_{-1}}s = e^{-2\phi}\pa_x^{-1}\(s e^{2\phi}\) \label{tm1gi}
\er
and
\begin{equation}
   \pa_{t_{-2}}r =-e^{2\phi}\partial_x^{-1}\left[\partial_x^{-1}\left( r e^{-2\phi}\right)-2re^{-2\phi}\partial_x^{-1}\left( r e^{-2\phi}\right) \partial_x^{-1}\left( s e^{2\phi}\right) \right], \label{tm2gir}
\end{equation}
\begin{equation}
   \pa_{t_{-2}}s=e^{-2\phi}\partial_x^{-1}\left[\partial_x^{-1}\left( s e^{2\phi}\right)+2se^{2\phi}\partial_x^{-1}\left( r e^{-2\phi}\right) \partial_x^{-1}\left( s e^{2\phi}\right) \right].\label{tm2gis}
\end{equation}
Equations (\ref{tm2gir}) and (\ref{tm2gis})  both admit zero and constant non-zero vacuum solutions.

In order to discuss the  vacuum structure of the GI  hierarchy  
let us define the Heisenberg algebra generated by
\br
\Sigma^ {(2n)} = \frac{1}{2} h^ {(n)} + b r_0E_{\a}^ {(n-1)} +b s_0 E_{-\a}^{(n)} -b r_0s_0 h^ {(n-1)}, \quad n=0,\pm 1, \pm 2, \cdots\label{sigma}
\er
{  For constant non-zero vacuum  values  we find for $A_x^ {vac}$ and $A_{t_{N}}^ {vac}$,
\br
A_x^ {vac }& =& \Sigma^{(2)} = \frac{1}{2}h^{(1)}+ r_0 E_\a^{(0)}
+s_0  E_{-\a}^{(1)} - r_0 s_0 h^ {(0)}, \nonu \\
 A_{t_2}^ {vac }& =&\Sigma^{(4)} , 
  \nonu \\
 A_{t_3}^ {vac }& =&\Sigma^ {(6)} +( r_0s_0)^2 \Sigma^ {(2)}, \nonu \\
 \vdots &=& \vdots, \nonu \\
 A_{t_N}^ {vac }& =&\sum_{j=1}^ {N} c_j (r_0s_0)^ {(N-j)} \Sigma^ {(2j)} , \quad N=1,2,3,\cdots
 \er

 For the negative grade sector  we find for the zero vacuum, i.e., $b=0$,
 \br
 A_{t_{-N}}^ {vac} = \frac{1}{2} h^ {(-N)}, \quad N=1, 2, \cdots
 \er
 and for the constant non-zero vacuum, i.e., $b=1$, we find for $t_{-2}$,
\br
 A_{t_{-2}}^ {vac} & = -\frac{1}{2r_0s_0} \Sigma^{(-2)}.
\er
We therefore expect  the following structure for the vacuum configuration
\br
A_{t_{-N}}^ {vac } =\sum_{i=0}^{N} c_i (r_0s_0)^{(-N+i) }\Sigma^{(-2i)}, \qquad  N =2,3, \cdots
\er
Observe  that from (\ref{atneg}), the graded structure of  $A_{t_{-N} }= D^ {(-N)}+ D^ {(-N+1)}+\cdots D^ {(-1)}$ in vacuum cannot  be fitted  in terms of  $\Sigma^ {(2n)} \sim X^{(2n)}+X^{(2n-1)}+X^{(2n-2)}, \quad X^ {(i)}\in \lie_i$ given in (\ref{sigma}) for some values of $n$, e.g., $n=-1$.  
This explains  the absence of $t_{{-1}}$ flow  for  constant non-zero vacuum.  This situation may be changed  by the introduction of $\tilde B$  as we shall now discuss.
 
  From equation (\ref{ax}) we find
that
$ A_x = \tilde  B \tilde A_x \tilde B^{-1} + \tilde B\pa_x \tilde B^{-1}  
=  (\Theta_<^ {-1} \eps^{(2)} \Theta_<)_{\ge}$ is given by (\ref{axgi}).  
   In terms of field variables $r$ and $s$ we  find (\ref{axrs}), 
  \begin{equation*}
  A_x=\frac{1}{2}h^{(1)}
 +r(t)E_{\alpha}^{(0)}+s(t)E_{-\alpha}^{(1)} -r(t) s(t) h^{(0)}.
   \end{equation*}
    We now choose $\tilde B =e^{-c\tilde \phi h^{(0)}}$ where $ \pa_x
    \tilde \phi = r(t)s(t)$ and gauge transform
    $A_x$ according to (\ref{gx}). It is convenient to define the $c$-dependent
    variables, {\footnote{Notice that unlikely the mKdV case, where
    the field variable $v(x,t_{\pm N})$ is independent of $c$, in this case the
    fields $R=R_c(x,t_{\pm N})$ and $S=S_c(x,t_{\pm N})$ are $c$-dependent and so are
    the flow equations. }}
\begin{equation}
 R_c(x,t_{\pm N})=r(t)e^{-2c\tilde \phi(t)}, \qquad S_c(x,t_{\pm N}) = s(t)e^{2c\tilde \phi(t)}.
 \label{RS}
\end{equation}
In this notation we find
  \begin{equation*}
   \tilde { A}_x
    =\frac{1}{2}h^{(1)}
    +R_c(t)E_{\alpha}^{(0)}+S_c(t)E_{-\alpha}^{(1)}-\left(1-c\right)R_c(t)S_c(t)h^{(0)}.
 \end{equation*} 
In order to simplify notation we shall drop  for  the moment 
the subscript $c$ from variables $R_c$ and $S_c$.  
Flow equations are  derived  by solving the zero curvature relation:
  \begin{equation*}
   \[ \mathcal{D}_{x}+\frac{1}{2}h^{(1)}-
   \left(1-c\right)R\left( t \right)S\left(t\right)h^{(0)}+R\left(t\right)
   E_{\alpha}^{(0)}+S\left(t\right)E_{-\alpha}^{(1)}, \;\; 
   \mathcal{D}_{t_{N}}+E^{(2N)}+\sum^{2N-1}_{i=0}D^{(i)} \] =0.
\end{equation*}
For  $N=2$ we find, 
\begin{equation}
\begin{aligned}
    A_{t_2}=&\frac{1}{2}h^{(2)}-RSh^{(1)}-\left (1-c\right )\left ( \left (1-4c\right )R^2S^2-\left (1-2c\right )b R_0^2S_0^2-SR_{x}+RS_{x} \right )h^{(0)}\nonu \\
    &+R E_{\alpha}^{(1)} +  SE_{-\alpha}^{(2)} -\left (2cR^2S + R_{x}  \right ) 
    E_{\alpha}^{(0)} -  \left (2cRS^2-S_{x}  \right
    )E_{-\alpha}^{(1)}\, ,
\end{aligned}
\end{equation}
 where $R_0$ and $S_0$ and the corresponding vacuum values of $R$ and $S$ while for  $N=3$,
\begin{equation}
\begin{aligned}
    A_{t_3}=&\frac{1}{2}h^{(3)}-RS h^{(2)}-\left (\left (1-4c\right ) R^2S^2-SR_{x}+RS_{x} \right )h^{(1)}\nonu \\
    &-\left (1-c\right )\left ( -2\left (1-6c^2\right )R^3S^3+4\left (1-2c\right )cb R_0^3S_0^3+6cRS^2R_{x}\right )h^{(0)}\nonu \\
    &-\left (1-c\right )\left ( -\left(6cR^2S+R_x \right)S_x+SR_{xx}+RS_{xx}\right )h^{(0)}\nonu \\
    &+R E_{\alpha}^{(2)} +  SE_{-\alpha}^{(3)} -\left (2cR^2S + R_{x}  \right ) E_{\alpha}^{(1)} -  \left (2cRS^2-S_{x}  \right )E_{-\alpha}^{(2)}\nonu \\
    &+\left ( 2R\left ( -\left (1-2c\left (1+c\right )\right )R^2S^2+3cSR_x-\left (1-c\right )RS_x\right )+R_{xx} \right )E_{\alpha}^{(0)}\nonu \\
    &+\left ( -2S\left (\left (1-2c\left (1+c\right )\right )R^2S^2-\left (1-c\right )SR_x+3cRS_x\right ) +S_{xx} \right )E_{-\alpha}^{(1)}
\end{aligned}
\end{equation}
and the corresponding flows equations are:
\begin{align}
&   \partial_{t_2}R=-\partial^2_xR+2\left (1-c\right)  \left ( 1-2c \right )  
R\left(R^2S^2-bR_0^2S_0^2 \right)+2\left ( 1-c \right
)R^2S_x-2c\partial_x\left(R^2S\right),\label{4.21a}
 \\
&   \partial_{t_2}S=\partial^2_xS-2 \left ( 1-c \right )
\left ( 1-2c \right ) S\left(R^2S^2-bR_0^2S_0^2 \right)+2\left ( 1-c \right )S^2R_x -
2c\partial_x\left ( RS^2 \right ), \label{4.20}\\
& \nonu \\
&   \partial_{t_3}R=\partial^3_xR-8c\left (  1-c \right )
\left ( 1-2c \right )R \left(R^3S^3 -bR_0^3S_0^3\right)
-6(1-c)(1-4c)R^2S^2R_x\nonu \\
   & -6(1-c)RR_xS_x-6c\partial_x\left[(1-2c)R^3S^2-RSR_x\right], \label{4.21}\\
&  \partial_{t_3}S=\partial^3_xS
  +8c\left (  1-c \right )\left ( 1-2c \right )S\left(R^3S^3 -bR_0^3S_0^3\right)
  -6(1-c)(1-4)R^2S^2S_x\nonu \\
  &+6(1-c)SR_xS_x-6c\partial_x\left[(1-2c)R^2S^3+RSS_x\right]. \label{4.22}
\end{align}
Notice that in this case $\tilde B E^{(2)} \tilde B^{-1} = E^{(2)}$ 
and the fields
$R(x,t_{\pm N})$ and $S(x,t_{\pm N})$ defined in (\ref{RS}), unlike in the mKdV case, 
change under
the gauge transformation (\ref{gx}) and (\ref{gt}). They are actually $c$-dependent 
and so are the corresponding flow equations, e.g.,
(\ref{4.20})-(\ref{4.22}), etc.  Since the hierarchy is invariant
under gauge transformation (\ref{gx}) and (\ref{gt}) and hence independent of $c$,
three interesting cases of the {\it same hierarchy} naturally arise
in terms of different gauge variant field variables defined in
(\ref{RS}) for $c=0$, $c=1/2$ and $c=1$.  For $c=0$ we find the
GI hierarchy already discussed earlier. When $c=1/2$ we get the CLL
hierarchy while  $c=0$ corresponds to KN hierarchy.  This
fact was already anticipated in ref. \cite{gui2} in terms of a change of
variables and in this paper we have established it on a more
fundamental basis  in terms of the g-RHB
decomposition (\ref{gr-h}). Interestingly, we  notice 
that the $t_2$ flows in \eqref{4.21a},\eqref{4.20} satisfy the
conservation law $(RS)_{t_2}=\lbrack (RS_x-SR_x)+(1-4c)
R^2S^2\rbrack_x$ that depends on the parameter $c$.

 We should like to  point out that  the CLL hierarchy obtained with 
 variables (\ref{RS})  for $c=1/2$ is the only hierarchy that admits
 non-zero constant vacuum solutions, i.e.
\begin{equation*} R=R_0, \qquad S=S_0.
 \end{equation*}
It is straightforward to verify  this 
since in  this limit   ($c=1/2$) equations (\ref{4.21a}) and (\ref{4.20}) become
\br
\pa_{t_2}R+\pa_x^2R + 2 R\; S \pa_x R = 0, \quad
\pa_{t_2}S-\pa_x^2S + 2 R\; S \pa_x S = 0,
\er
and
\br
&\pa_{t_3}R-\pa_x^3R - 3 R^2\; S^2 \pa_x R-3SR^2_x -3 RS R_{xx} = 0, \nonu \\
&\pa_{t_3}S+\pa_x^3S - 3 R^2\; S^2 \pa_x S  + 3RS^2_x +3 RSS_{xx}= 0.
 \er
 
Next, we derive the negative sub-hierarchy. Consider
 \br 
 A_{t_{-1}} =\frac12 h^ {(-1)} +\bar Re^ {2(1-c)
 \pa_x^ {-1}(R S)}E_{\a}^ {(-1)} -\bar S
 e^ {-2(1-c)\pa_x^ {-1}(R S)}E_{-\a}^ {(0)}+ c\bar R \bar S h^
 {(0)} \, ,
 \er
     where 
  \begin{equation}
  \bar R = \pa_x^ {-1} \(Re^ {-2(1-c)\pa_x^ {-1}(R S)}\), 
  \qquad  \bar S= \pa_x^ {-1} \(Se^ {2(1-c)\pa_x^ {-1}(R S)}\) \, .\label{tilde}
  \end{equation}
 It follows that
 \begin{equation}
  \pa_{t_{-1}} R = \bar R e^ {2(1-c)\pa_x^ {-1}(RS)}-2c R \bar R \bar
  S, \qquad \pa_{t_{-1}} S = \bar S e^ {-2(1-c)\pa_x^ {-1}(RS)}+2c S
  \bar R \bar S \, .
  \label{laf1}
  \end{equation}
  For $c=1$,  the above equation can be put in a local form by defining  
  new variables,
  \begin{equation}
   \bar R = \pa_x^ {-1} \(R \) \equiv \Phi_R, \qquad  
  \bar S= \pa_x^ {-1} \(S\) \equiv \Phi_S \, ,\label{tildec1}
  \end{equation}
 and equation (\ref{laf1})  becomes \cite{gui2}, 
 \begin{equation}
 \pa_{t_{-1}}\pa_x\Phi_R = \Phi_R -2 \pa_x \Phi_R( \Phi_R \Phi_S) , 
 \qquad \pa_{t_{-1}}\pa_x\Phi_S = \Phi_S -2 \pa_x \Phi_S (\Phi_R \Phi_S).
  \label{phirs}
 \end{equation}
 A class of higher grading relativistic integrable  models  with similar 
 structure has been considered in \cite{LAF} in connection with reductions
 of WZWN model.
  It is clear that  equation (\ref{phirs}) only admits zero vacuum solution. 
  
A condition for non-zero vacuum solution can be found by considering $R=R_0, 
\quad S=S_0 = $ constants and obtaining from relation \eqref{tilde}:
\begin{align}
\bar R_{vac}  &=R_0  \int^x  e^{-2(1-c)R_0 S_0\;  y}\;\; dy 
= \frac{R_0}{-2(1-c)R_0S_0}e^ {-2(1-c)R_0 S_0 x}, \label{RSvacR} \\
 \bar S_{vac}  &= S_0 \int^x  e^ {2(1-c)R_0 S_0\;  y}\;\; dy 
 = \frac{S_0}{2(1-c)R_0S_0}e^{2(1-c)R_0 S_0 x}\, .  \label{RSvacS}
\end{align}
 Substituting equations \eqref{RSvacR} and \eqref{RSvacS}
 in both of equations (\ref{laf1}) we find  the following 
 condition for $c$: 
 \br
\frac{1-2c}{2(1-c)^2}=0, \label{c}
\er
  which implies that the only constant  non-zero  vacuum solution  for $t_{-1}$-flow
  occurs for $c=1/2$.
    
For the next negative flow for $t= t_{-2}$ we find 
\begin{equation*}\begin{split}
\pa_{t_{-2}} R &=-e^{2(1-c)\partial_x^{-1}\left( R S \right)}
\partial_x^{-1}\left( \bar R - \bar S \partial_x \bar R^2
\right)\\
    &+2cR \left[ \bar R^2\bar S^2 -\bar R\partial_x^{-1}
    \left( \bar S + \bar R \partial_x \bar S^2 \right)+
    \bar S\partial_x^{-1}\left( \bar R - \bar S \partial_x 
    \bar R^2 \right) \right], \\
 \pa_{t_{-2}} S &= e^{-2(1-c)\partial_x^{-1}\left(R S \right)}
 \partial_x^{-1}\left( \bar S + \bar R \partial_x \bar S^2 \right)\\
    &-2cS\left[ \bar R^2\bar S^2 -\bar R\partial_x^{-1}
    \left( \bar S + \bar R \partial_x \bar S^2 \right)+
    \bar S\partial_x^{-1}\left( \bar R - \bar S \partial_x 
    \bar R^2 \right) \right].
 \end{split}\end{equation*}
 Again,  non-zero  vacuum solution (\ref{RSvacR})and (\ref{RSvacS}) 
 implies equation (\ref{c}) indicating  that
  $c=1/2$ is indeed  a special value.  
 We conclude  that the CLL hierarchy is the only one
  of the $a=2$  models that admits constant non-zero vacuum  solution.
  In fact in  reference \cite{thais-paper} we  have constructed 
  explicit solutions for the CLL hierarchy  in terms of vertex operators.  
  These are classified   in terms of ``two flavor" vertices, namely 
  $ V_{\pm}(z)$.  Non-trivial solutions for fields $ R$ and $ S$ 
  are obtained from mixed  powers of products of these two vertices.    
  Another class of solutions, keeping one of the fields  constant is 
  obtained by considering powers of one of the vertices.   
  The latter  gives rise to   Burgers  hierarchy where one of the CLL fields 
  is set to  identity as we discuss in the next sub-section.
 
 \subsection{The Burgers Hierarchy}
 
 Having settled the fact that the CLL hierarchy  admits  non-zero
 constant vacuum solutions, we  are now  in position  to explore a particular  set of solutions obtained by setting one of the fields ($R$ or $S$)  to a constant (normalized to one).   This gives rise to  the Burguers hierarchy  as a sub-hierarchy with
$R=1$ and $S= w$ for $c=1/2$ {\footnote{ In ref. \cite{thais-paper} we
  develop an explicit construction of  such class of solutions using 
  the dressing method and classify them  by judicious choice of  products of 
  vertex operators }}.  The  first two flow equations are
\begin{equation*}\begin{split} 
 \pa_{t_2} w &= \pa_x^2 w -2 w \pa_x w,  \\
  \pa_{t_3} w &= \pa_x^3 w +3 w^2  \pa_x w - 3\pa_x (w \pa_x w), 
\end{split}\end{equation*}
 and for the negative sector we find
 \begin{equation*}\begin{split} 
 \pa_{t_{-1}} w &= 1 +  w e^{\pa_x^{-1} w} \pa_x^{-1} \( e^{-\pa_x
 ^{-1}w}\) \, ,
  \\
 \pa_{t_{-2}} w &=  e^{\pa_x^{-1} w} \pa_x^{-1} \( e^{-\pa_x ^{-1}w}\) +
 w e^{\pa_x^{-1} w} \pa_x^{-2} \( e^{-\pa_x ^{-1}w}\) \, , \\
 \pa_{t_{-3}} w &=  e^{\pa_x^{-1} w} \pa_x^{-2} \( e^{-\pa_x ^{-1}w}\) +
 w e^{\pa_x^{-1} w} \pa_x^{-3} \( e^{-\pa_x ^{-1}w}\).
\end{split}\end{equation*}
 The positive and negative flows for the sub-hierarchies can be written in closed form as
\begin{equation*}
    \pa_{t_{N}}w=-\,\pa_x \(e^{\pa_x^{-1} w}
 \pa_x^{N} \( e^{-\pa_x ^{-1}w}\)\),
\end{equation*}
for $ N=1, 2, 3, \ldots$. We observe that this reduction procedure applied on the CLL model reproduces a general closed formula  for all positive flows of the Burgers hierarchy, see for example  \cite{kudria}. Here the reduction process yields
\begin{equation*}
    \pa_{t_{-N}}w=\,\pa_x \(e^{\pa_x^{-1} w}
 \pa_x^{-N} \( e^{-\pa_x ^{-1}w}\)\),
 \end{equation*}
for $ N=1, 2, 3, \ldots$,  extending the definition of the Burgers hierarchy to include negative flows of the same structure.

\section{A Novel Model.  Higher grade hierarchy with $a=3$ with
principal gradation and $\hat \lie=\hat{sl}_2$}
\label{section:higher}
Here we  consider $\hat \lie=\hat{sl}_2$ with the principal gradation, but
set
the parameter $a$ to be equal to $a=3$ (higher $a$ then we have
encountered up to now). Accordingly, the semi-simple element
inducing the corresponding hierarchy is then equal to
$  E^{(3)}=\frac{1}{2}(E_{\alpha}^{(1)}+E_{-\alpha}^{(2)})$.
From equation \eqref{ax}, we parametrize $\theta^{(-3)}=\chi_{-3} E_\a^{(-2)} +\psi_{-3}E_{-\a}^{(-1)}, \quad \theta^{(-2)}=\phi_{-2}h^{(-1)}. \quad \theta^{(-1)}=\chi_{-1}E_\a^{(-1)} +\psi_{-1}E_{-\a}^{(0)}$
and obtain 
\begin{equation}
\begin{aligned}
    \label{Ax42}
    \tilde{A}_{x,c}=&\, \frac{1}{2}\left(e^{-2c\,\tilde\phi}E_{\alpha}^{(1)}+e^{2c\,\tilde \phi}E_{-\alpha}^{(2)}\right)+\frac{1}{2}\left( \psi_{-1}-\chi_{-1}\right)h^{(1)}+\\
    &-\left(\phi_{-2}+\frac{1}{2}\chi_{-1}\left( \psi_{-1}-\chi_{-1}\right)\right)e^{-2c\,\tilde \phi}E_{\alpha}^{(0)}+\\
    &+\left( \phi_{-2}-\frac{1}{2}\psi_{-1}\left( \psi_{-1}-\chi_{-1}\right)\right)\,e^{2c\,\tilde \phi}E_{-\alpha}^{(1)}+\\
    &+\left(\frac{1}{3}\chi_{-1}\psi_{-1}\left( \psi_{-1}-\chi_{-1}\right)-\frac{1}{2}\phi_{-2}\left( \psi_{-1}+\chi_{-1}\right)\right) h^{(0)}+\\
    &+\frac{1}{2}\left( \psi_{-3}-\chi_{-3}\right) h^{(0)}+c\,\tilde\phi_x h^{(0)}.
\end{aligned}
\end{equation}
Setting
$v_1 = \frac{1}{2}(\psi_{-1}-\chi_{-1})$, $v_2=\phi_{-2}-v_1\psi_{-1}$
and 
$ v_{3}= \frac{1}{2}(\psi_{-3}-\chi_{-3})-v_2 \psi_{-1} - \frac{1}{3}v_1(v_1+\psi_{-1}) $, where $v_1=v_1(t)$, $v_2=v_2(t)$ and $v_3=v_3(t)$, results
in 

\br
\label{Ax44}
    \tilde{A}_{x}&=&\,\frac{1}{2}\left(e^{-2c\,\tilde\phi}E_{\alpha}^{(1)}+e^{2c\,\tilde\phi}E_{-\alpha}^{(2)}\right)+v_1h^{(1)}-\left(2v_{1}^2+v_{2}\right)e^{-2c\,\tilde\phi}E_{\alpha}^{(0)}+v_{2}e^{2c\,\tilde \phi}E_{-\alpha}^{(1)}
    -\left(1-c\right)\tilde \phi_x h^{(0)}.\nonu
\er
Let   { $\tilde \phi_x = -v_1v_2-v_3$}.
}
We get for the lowest flow
\begin{equation}
\begin{aligned}
\label{v1}
    v_{1,t_3}=&\,v_{1,3x}-6\left ( v_1^2+v_2 \right )\left (v_{1,x}^2+\left ( v_2^2-2v_1v_3 \right )^2\right )-6\partial_x\left ( v_1v_3^2 \right )  +\\
    &-6v_3\partial_x\left ( v_1^4+2v_1^2v_2 \right )+
    2v_2^2\partial_x\left ( v_1^3+3v_3+3v_1v_2 \right ) \, ,
\end{aligned}
\end{equation}
\begin{equation}
\begin{aligned}
\label{v2}
    v_{2,t_3}=&\,v_{2,3x}+6\left ( v_1v_2-v_3 \right )\left (v_{1,x}^2+\left ( v_2^2-2v_1v_3 \right )^2\right )-6v_3^2v_{2,x}+\\
    &-6\partial_x\left( v_1^2v_2^3+\left ( v_2^2-2v_1v_3 \right )v_{1,x}\right)-12v_1v_2\partial_x\left( v_2v_3\right)+\\
    &+4v_1^2\partial_x\left(
    v_2^3+3v_3^2\right)+6\left(v_{1,xx}+4v_1^2v_2v_3\right)v_{1,x} \, ,
\end{aligned}
\end{equation}
\begin{equation}
\begin{aligned}
\label{v3}
    v_{3,t_3}=&\,v_{3,3x}-6\left ( v_2^2+v_1v_3 \right )\left (v_{1,x}^2-\left ( v_2^2-2v_1v_3 \right )^2\right )-6\left (v_2^3+v_3^2\right )v_{3,x}+\\
    &-6\partial_x\left( -4v_1^3v_3^2+3v_1^2v_2^2v_3-2v_1^2v_3v_{1,x}+\left ( v_2^2-2v_1v_3 \right )v_{2,x}\right)+\\
    &+6v_1v_2\partial_x\left( v_2^3-v_2v_{1,x}\right)-6v_{1,x}\left(v_1v_2v_{2,x}-\partial_x\left( v_1v_{1,x}\right)\right)+\\
    &+24v_1v_3\left( v_{1,x}^2-v_1^2v_{3,x}\right)+3\partial_x\left(
    v_{1,x}v_{2,x}\right) \, .
\end{aligned}
\end{equation}

The above flows are the lowest evolution equations of a novel 
higher grading hierarchy characterized by $a=3$.
To reduce these equations to the $a=1$ mKdV hierarchy one sets 
$v_1=0$, $v_2=0$ for which the flow equations \eqref{v1} and
\eqref{v2} identically vanish. The remaining variable $v_3=v$ satisfies the mKdV
$t_3$-flow \eqref{mkdv} that agrees with equation \eqref{v3} in
this limit.

\section{ Conclusions and Further Developments}

In this paper we have proposed a unified universal construction of a
class of integrable hierarchies based upon a generalization of
Riemann-Hilbert-Birkhoff decomposition, which incorporates  higher graded structures and different types of vacuum. The key
ingredient is the decomposition of the affine algebra $\hat {sl}_2 $
into graded subspaces according to the principal gradation and a
choice of semi-simple  grade $a$ constant generator
$E=E^{(a)}$.  The construction is systematic and generates a series of
flow equations associated to positive and negative graded Lie
algebraic elements.  As a basic $a=1$ example our construction generates the
well-known mKdV hierarchy which naturally splits into two classes of
solutions, those constructed from zero vacuum, for positive and
negative odd graded flows and those build up from non-zero constant
vacuum solutions associated to positive odd and negative even flows.
The second example concerns a grade 2 construction yielding three
known hierarchies namely, KN, CLL and
GI hierarchies. In fact our construction indicates that
they differ by a change of variables induced by an ambiguity in
parameterizing the zero graded sector in g-RHB decomposition.
Although these three hierarchies are equivalent, we show that the CLL
hierarchy stands out in the sense that it is the only one admitting non-zero
constant vacuum solution for all flows.  Such peculiar property allows
the reduction of the CLL to the Burgers hierarchy where one of the CLL
fields is set to constant.  In fact, the construction of soliton
solutions for the CLL hierarchy \cite{thais-paper} in terms of
deformed vertex operators shows that this is indeed the case.

We should point out that the construction developed in this paper,
apart from providing an advantage of unifying  several 
known hierarchies 
within the same structure, also paves the way
towards systematic construction of higher graded hierarchies
as exemplified by $t_3$-flow of the $a=3$ model of
section \ref{section:higher}.  Of course many new hierarchies 
may be constructed within
the same universal framework considering the same $\hat {sl}_2$
affine algebra and principal gradation for $a=4$ and higher values of  $a$.  An interesting study of such
framework involving homogeneous gradation shows connection between
AKNS, WKI and other hierarchies that will be developed elsewhere \cite{thais-thesis}.

Generalizations of this construction to higher rank affine algebras,
as well as, inclusion of other gradations may uncover
 interesting new and relationships among integrable hierarchies.

{\large \bf Acknowledgments}

	JFG and AHZ thank CNPq and FAPESP for support. TCS thanks CNPq for financial support.

 \end{document}